\documentstyle[aps, twocolumn, epsfig]{revtex}
\begin{document}
\draft
\title{Phase relaxation of Faraday surface waves}
\author{A.V.~Kityk$^{1,4}$, C. Wagner$^2$, K.~Knorr$^1$, H.~W.~M\"{u}ller$^{3,1}$}
\address {$^1$Fakult\"{a}t f\"{u}r Physik und Elektrotechnik, Universit\"at des Saarlandes, 66041
Saarbr\"ucken, Germany\\
$^2$Laboratoire de Physique Statistique, Ecole Normale
Sup\'{e}rieure, 24 rue
Lhomond, 75231 Paris Cedex 05, France\\
$^3$Max-Planck-Institut f\"{u}r Polymerforschung, Ackermannweg 10, 55128 Mainz, Germany \\
$^{4}$Czestochowa Technical University, Electrical Eng. Dep., Al.
Armii Krajowej 17, PL-42200 Czestochowa, Poland } \maketitle
\date{\today}

\begin{abstract}
Surface waves on a liquid air interface excited by a vertical
vibration of a fluid layer (Faraday waves) are employed to
investigate the phase relaxation of ideally ordered patterns. By
means of a combined frequency-amplitude modulation of the
excitation signal a periodic expansion and dilatation of a square
wave pattern is generated, the dynamics of which is well
described by a Debye relaxator. By comparison with the results of
a linear theory it is shown that this practice allows a precise
measurement of the phase diffusion constant.

\end{abstract}
\pacs{47.54.+r, 47.20.Ma}

Our understanding of spatio-temporal pattern formation in
non-equilibrium fluid systems has greatly benefitted
\cite{cross93} from recent quantitative experiments in
combination with the development of new theoretical concepts. One
of them is the so-called {\em amplitude equation} approach
\cite{newell69}, which is based on the linear instability of a
homogeneous state and leads naturally to a classification of
patterns in terms of characteristic wave numbers and frequencies.
A different but equally universal description, the {\em phase
dynamics} \cite{pomeau79}, applies to situations where a periodic
spatial pattern experiences long wavelength phase modulations.
This approach, originally introduced in the context of thermal
convection, has proven to be useful to understand the stability
and the relaxation of periodic patterns, wave number selection,
and defect dynamics. In many paradigmatic pattern forming systems
such as thermal convection in a fluid layer heated from below
(Rayleigh B\'{e}nard convection, RBC) or the formation of azimuthal
vortices in the gap between two rotating cylinders (Taylor
Couette flow, TCF) the dominating wave number is dictated by the
geometry and thus inconvenient to be changed in a given
experimental setup (for instance by a mechanical ramp of the
layer thickness \cite{kramer82}).

Faraday waves are surface waves on the interface between two
immiscible fluids, excited by a vertical vibration of the
container. Beyond a sufficiently large excitation amplitude the
plane interface undergoes an instability (Faraday instability)
and standing surface waves appear, oscillating with a frequency
one half of the drive. This type of parametric wave instability is
attractive as the wavelength of the pattern is {\em dispersion}
rather than {\em geometry} controlled. Just by varying the drive
frequency the wave number can be tuned in a wide range. In that
sense the Faraday setup is well suited for the study of phase
dynamics.

Nevertheless recent research activity in this field was mainly
dedicated to the exploration of the processes underlying the
selection of patterns with a fixed wavelength. Faraday
\cite{faraday31} was the first to provide a quantitative study of
this system, revealing that a sinusoidal vibration may induce a
periodic array of squares. Later on, more complicated patterns
with up to a 12-fold rotational symmetry (quasi-periodic
structures) have been observed \cite{douady90,muller98}. Here the
amplitude equation technique contributed considerably to unfold
the governing spatio-temporal resonance mechanisms. Applied to a
set of modes ${\bf k}_i$ with different orientations but fixed
wavelength, $|{\bf k}_i|=k$ the resulting set of Landau equations
lead to a semi-quantitative understanding \cite{milner91} of
pattern selection in this system. Motivated by these advances the
idea came up to apply more complicated drive signals composed of
two or more commensurable frequencies \cite{edwards93}. That way
the simultaneous excitation of distinct wavelengths gave rise to
novel surface patterns in form of superlattices
\cite{edwards93,kudrolli98,arbell00}. Only recently the phase
information carried by the participating modes was found to have
a crucial influence on the visual appearance of the convection
structures \cite{wagner01}.

In comparison to other classical pattern forming systems such as
RBC or TCF, the phase dynamics in the Faraday system is much less
explored. In usual Faraday experiments the drive frequency (or
frequency composition including relative amplitudes) is held
fixed while the overall drive amplitude is ramped in order to
record the bifurcation sequence of appearing structures.  To our
knowledge none of the previous investigations used the excitation
frequency $\omega$ as the primary control parameter rather than
the drive amplitude $a$. That way it is particularly simple to
impose phase perturbations on ordered patterns and to study their
relaxation dynamics. Moreover, doing phase dynamics on the
Faraday system has the additional advantage of rather quick
relaxation times, which in typical setups are one and two order
of magnitude faster than for instance in RBC.

The present paper reports a systematic investigation of phase
relaxation on Faraday surface waves. Our study is focused on the
relaxational dynamics of an ideal surface pattern with a square
tesselation. By evaluating the relaxation time of the pattern in
response to small changes of the frequency, the phase diffusion
coefficient has been measured. The experimental results are found
to be in good agreement with the findings of the linear theory
evaluated for a system of infinite lateral extension.

\begin{figure}
\epsfig{file=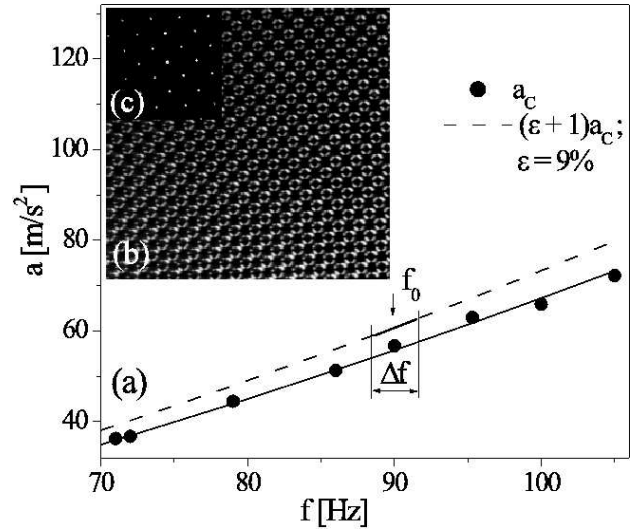, angle=0, width=0.95\linewidth}
\vspace{0.5cm} \caption{(a) Symbols denote the critical
acceleration amplitude $a_c$ as measured at the onset of the
Faraday instability at constant drive frequency $\omega=2 \pi f$.
The solid line indicates the theoretical prediction for the
material parameters given. Along the dashed line the reduced drive
amplitude is $\varepsilon=9\%$. The phase relaxation experiments
were performed by imposing simultaneous small changes of $f$ and
$a$, as indicated for an example by the solid bar. (b) and (c)
show a square pattern and the associated 2-dimensional power
spectrum as observed during the modulation experiments.}
\label{fig1}
\end{figure}

The experimental setup consists of a black cylindrical container
built out of anodized aluminium, and filled to a height $h$ of
$4.2$ mm with a silicone oil (kinematic viscosity $\nu=21.4
\times 10^{-6}$ m$^2$/s, density $\rho=949$ kg/m$^3$, surface
tension $\sigma=17.3 \times 10^{-3}$ N/m).  In order to study
finite size effects, we used three different containers, with
inner diameters $L_1=265$ mm, $L_2=185$ mm, and $L_3=125$ mm.  A
glass plate covering the container was used to prevent
evaporation, pollution and temperature fluctuations of the liquid.
Furthermore, to avoid uncontrolled changes of the viscosity,
density, and surface tension of the liquid, all the measurements
have been performed at a constant temperature of $30\pm
0.1^\circ$C.  The Faraday waves were excited by an
electromagnetic shaker  vibrating vertically with an acceleration
allowing for simultaneous amplitude {\em and} frequency
modulations in the form $a(t)\cos{\omega(t)}$. The corresponding
input signal was produced by a waveform generator via a
D/A-converter. The instantaneous acceleration was measured by a
piezoelectric sensor. In a preparatory experiment undertaken with
a sinusoidal (i.e. unmodulated) drive $a \cos{\omega t}$ the
critical acceleration amplitude $a_c(\omega)$ for the onset of
the Faraday instability was determined by visual inspection of
the interface while quasi-statically ramping $a$ at fixed
$\omega=2 \pi f$ (see Fig.~\ref{fig1}). Throughout the
investigated frequency interval $70$ Hz$<f<110$ Hz the surface
patterns, which appear at a supercritical drive of less than
about $1.1 \times a_c$, always consisted of an ordered square
wave pattern, which -- after some healing time -- was free of
defects (Fig.~\ref{fig1}b). In order to study the dynamics of
phase-perturbed patterns we have carried out measurements of the
average wave number $k(t)$ of the Faraday pattern in response to
small changes of the drive frequency $\omega(t)$ around a mean
value $\omega_0$. The $\omega$-modulation has been accomplished
in two different ways: (i) By discontinuous jumps (back and
forth) between frequencies $\omega_0-\Delta \omega/2$ and
$\omega_0+\Delta \omega/2$ with a repetition period $T$ between
$100$-$300$s, sufficiently large for the pattern to relax. (ii)
By a sinusoidal modulation of the drive frequency according to
$\omega(t)=\omega_0 + \Delta \omega \sin{(\Omega t)}$, with
$T=1/F=2 \pi/\Omega$ between $2$ and $1000$s. (within the
frequency range of our study, the response time of the shaker to
small changes of the drive frequency is less than $10$ ms and
thus negligible). In both cases the vibration amplitude $a(t)$
was co-modulated in such a way that the instantaneous
supercritical drive $\varepsilon=a(t)/a_c[\omega(t)]-1$ remained
constant (see Fig.~\ref{fig1}a). The bandwidth $\Delta \omega$ of
the modulation needed to be confined to a few Hz in order to avoid
the occurrence of dislocation-type defects. Under those
conditions the square pattern remained practically ideal without
perturbations (see Fig.~\ref{fig1}b), just expanding and
contracting (breathing) in a homogeneous manner with the
modulation period  $T$. To obtain the temporal wave number
dependence a full frame CCD camera surrounded by a set of 4
incandescent lamps was mounted some distance above the container.
About $100$ pictures of the light reflected from the surface were
taken at consecutive instances of maximum surface excursion from
which the spatially averaged wave number $k(t)$ was extracted by
evaluating the position of the principal peaks in a 2-dimensional
FFT.

\begin{figure}
\epsfig{file=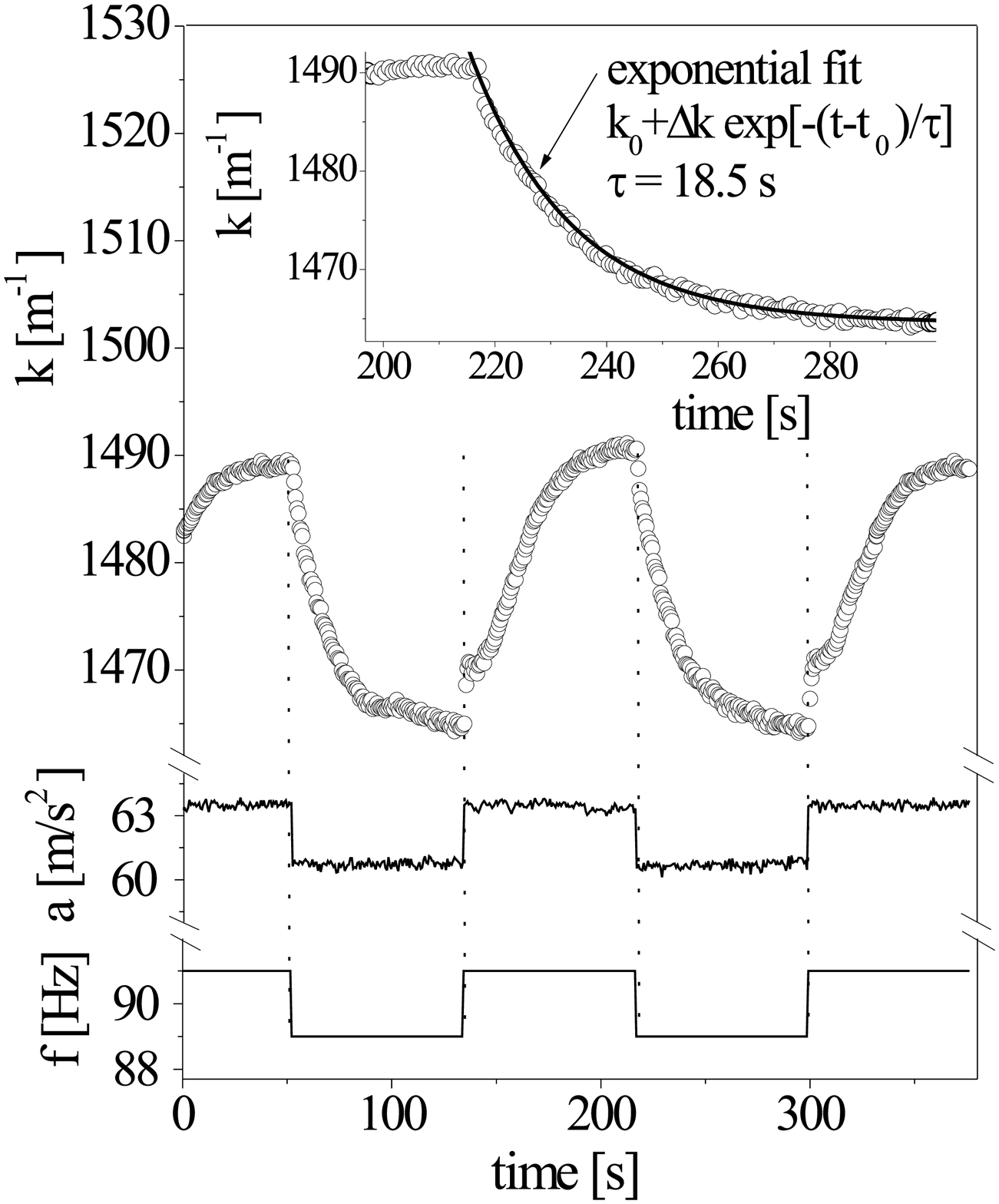, angle=0, width=0.9\linewidth}
\vspace{0.5cm} \caption{Temporal decay of the wave number $k(t)$
in response to a step-like [type (i)] change of the drive
frequency $f$. Also the drive amplitude $a$ was changed to keep
the reduced amplitude $\varepsilon=a(t)/a_c[\omega(t)]-1$
constant. The decay  of the wave number $k(t)$ could be fitted by
an exponential. (Parameters: $L=0.265$ m, $\varepsilon = 9 \%$,
$f= 90$ Hz) }\label{fig2}
\end{figure}

\begin{figure}
\epsfig{file=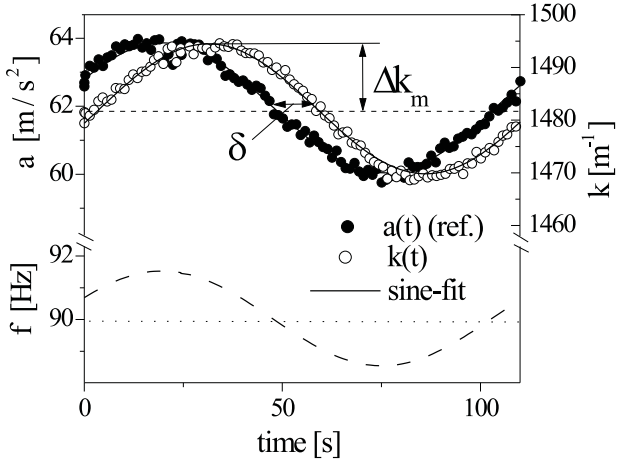, angle=0, width=0.94\linewidth}
\vspace{0.5cm} \caption{Relaxational dynamics of the wave number
$k(t)$ in response to a sinusoidal [type(ii)] modulation of the
drive frequency. The drive amplitude $a$ was co-modulated to keep
the reduced amplitude $\varepsilon$ constant. The Faraday pattern
responds with a sinusoidal variation of its instantaneous wave
number $k(t)$ with an amplitude $\Delta k_m$ and a temporal phase
lag $\delta$ (Parameters: $L=0.265$ m, $\varepsilon = 9 \%$, $f=
90 $ Hz).} \label{fig3}
\end{figure}

\begin{figure}
\epsfig{file=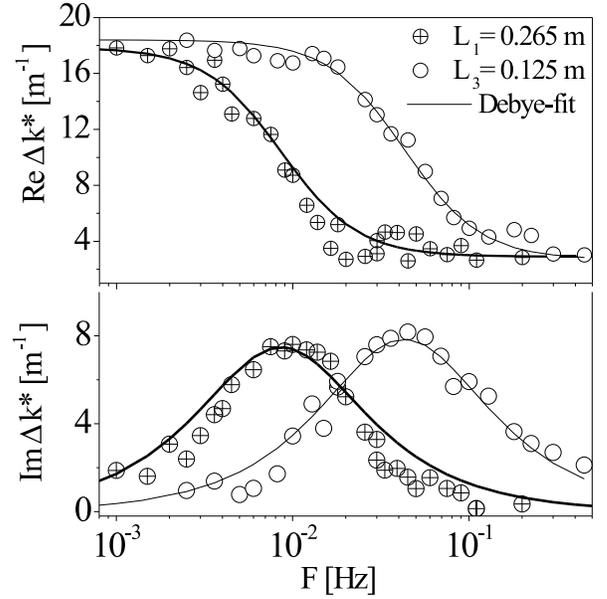, angle=0, width=0.90\linewidth}
\vspace{0.5cm} \caption{Frequency ''dispersion'' of the real and
imaginary parts of $\Delta k^*$ as obtained from modulation
experiments of type (ii) for the two different container diameters
$L_1=265$ mm and $L_3=125$ mm ($f= 90$ Hz, $k_0=1480$ m$^{-1}$,
$\varepsilon = 9 \%$). Solid curves are fits according to the
Debye relaxator }\label{fig4}
\end{figure}

\begin{figure}
\epsfig{file=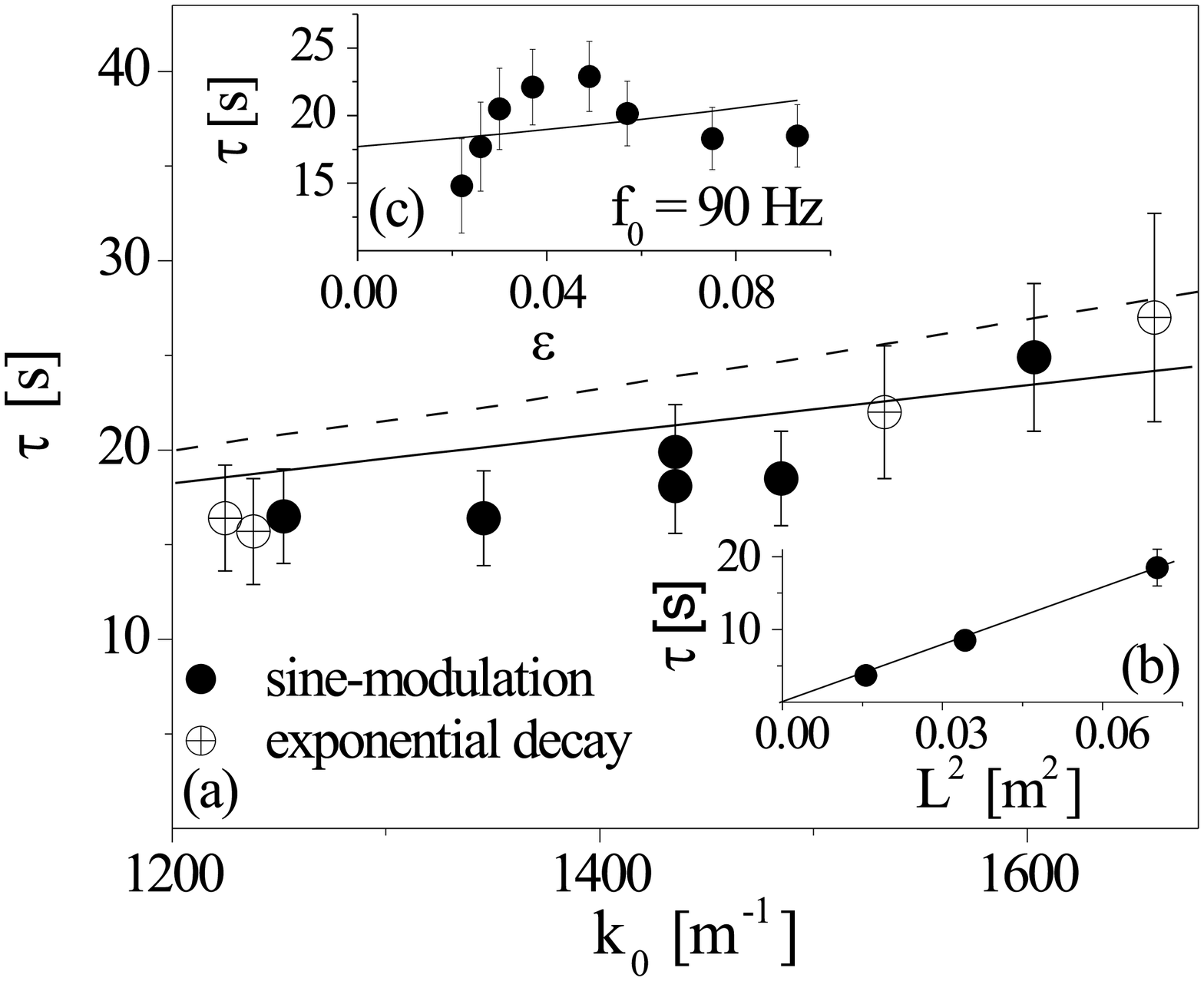, angle=0, width=0.90\linewidth}
\vspace{0.7cm} \caption{The characteristic relaxation time $\tau$
 as a function of (a) the mean wave
number $k_0$, (b) the container size $L$, and (c) the reduced
drive strength $\varepsilon$. The dashed line shows the
prediction for infinite depth capillary waves [Eq.~(\ref{tau})],
while the solid line is based on a numerical evaluation for the
(more realistic) case of finite depth gravity-capillary waves.
(Parameters: $L=0.265$ m $\varepsilon = 9 \%$, $f=90$ Hz)
}\label{fig5}
\end{figure}

The wave number $k(t)$ followed the modulation in a relaxational
manner. For the discontinuous modulation (i) this is directly
apparent from Fig.~\ref{fig2}. Here the relaxation time $\tau$
has been derived by fitting the exponential decay of the data. In
the type (ii) experiment $k(t)$ oscillates around a mean value
$k_0$ with an amplitude $\Delta k_m$ and a temporal phase lag
$\delta$ (see Fig.~\ref{fig3}). Introducing the complex wave
number increment $\Delta k^\star={\rm Re} [\Delta k^\star] + i \,
{\rm  Im}[\Delta k^\star]=\Delta k_m \, e^{i \delta}$, its real
and imaginary parts are plotted in Fig.~\ref{fig4} as a function
of the modulation frequency $F$. The solid curves of this figure
are fits of a linear Debye relaxator, where $\Delta k^*=\Delta
k(\Omega =0)/[1+i\Omega \tau]$. Fig.~\ref{fig5} shows the
relaxation time $\tau$, as obtained from both types of
experiments, as a function of the mean wave number $k_0$, the
container diameter $L$ and the reduced drive strength
$\varepsilon$. A dependence on the (small) modulation amplitude
$\Delta \omega$ could not be detected. The experimental data
reveal a linear increase with the wave number $k_0$ and a
proportionality to the square of the container size $L$.

The dependence of the relaxation time $\tau$ on, respectively,
$k_0$, $L$, and $\varepsilon$ can be understood in terms of the
phase diffusion approach \cite{pomeau79}. Here one takes
advantage of the fact that local disturbances of the elevation
amplitude die out rapidly, while long-wavelength phase
perturbation survive on a much longer (diffusive) time scale. To
streamline the arguments and to work out the basic physics we
consider a 1-dimensional surface elevation profile in form of
stripes as given by $\zeta(x,t) \propto [e^{i k_0 x +
\varphi(x,t)} + c.c.]\cos{(\omega t/2)} $. Here $\partial_x
\varphi=\Delta k$ describes the spatio-temporal variation of the
local wave number around the underlying base pattern with the
wave number $k_0$. Following the phase diffusion approach the
phase $\varphi$ obeys a diffusion equation of the form
$\partial_t \varphi = D_\parallel
\partial_{xx} \varphi $ with the diffusion constant (valid close
above onset) of the form
\begin{equation}\label{diffusion}
D_\parallel=\frac{\xi_0^2}{\tau_0} \, \frac{\varepsilon - 3
\xi_0^2 (k_0-k_c)^2}{\varepsilon - \xi_0^2 (k_0-k_c)^2}.
\end{equation}
The coefficients $\tau_0^{-1}=\partial \lambda /
\partial \varepsilon|_{k=k_c,\varepsilon=0}$ and
$\xi_0^2=-\tau_0/2 (\partial^2 \lambda /
\partial k^2)|_{k=k_c,\varepsilon=0}$ are given in terms of the linear
growth rate $\lambda=\lambda(\varepsilon,k)$ at which plane wave
perturbations grow out of the plane undeformed interface, when the
Faraday instability sets in. Here $k_c$ is the wave number at
onset of the Faraday instability. For weakly damped ($\nu
k_0^2/\omega \ll1$) capillary waves [$k_0 \simeq
(\rho/\sigma)^{1/3}\, (\omega/2)^{2/3}$] on a deep ($kh\gg 1$)
fluid layer (reasonable approximations in our experiment) one
obtains approximately $\tau_0^{-1} \simeq 2 \nu k_0^2$ and
$\xi_0^2=(1/2) [9 \sigma/(16 \nu^2 \rho)] k_0^{-3}$. By
decomposition of the phase perturbations into a set of discrete
Fourier modes compatible with the finite container dimension,
$\varphi = \sum_{n=1}^\infty a_n \sin{(n \pi/L)}$, the mode $n=1$
has the slowest decay time
\begin{equation}\label{tau}
\tau=\frac{L^2}{\pi^2} D_\parallel^{-1}= \frac{\nu \rho}{\sigma}
\,  k_0 \, \frac{16 L^2}{9 \pi^2 } \, \frac{\varepsilon -
\xi_0^2(k_0-k_c)^2}{\varepsilon - 3 \xi_0^2 (k_0-k_c)^2}
\end{equation}
and thus determines the relaxation time of the wave number of the
experiment. The dashed line in Fig.~\ref{fig5} is the prediction
according to Eq.~(\ref{tau}). For a quantitatively more reliable
check we also evaluated the coefficients $\tau_0$ and $\xi_0^2$
numerically from a linear analysis, which takes into account the
finite filling level as well as gravitational contributions to
the wave dispersion. The respective result is shown by the solid
line in Fig.~\ref{fig5}a. Furthermore, the predicted quadratic
dependence of $\tau$ on the container dimension $L$ is verified by
Fig.~\ref{fig5}b. The theoretical prediction Eq.~(\ref{tau}) also
implies a slight dependence of $\tau$ on the drive amplitude
$\varepsilon$. However, checking for this feature requires to
account for the fact that also the mean wave number $k_0$ is
affected by the drive strength. We deduced the empiric dependence
$k_0/k_c=1-\beta \varepsilon$ with $\beta=0.264$ from a control
run where $\varepsilon$ was slowly ramped at fixed $f$. Inserting
this result into the last term on the right hand side of
Eq.~(\ref{tau}) leads to the following expression
\begin{equation}\label{correction}
\frac{\tau(\varepsilon)}{\tau(\varepsilon=0)}=\frac{1- \xi_0^2
k_c^2 \beta^2 \varepsilon}{1- 3 \xi_0^2 k_c^2 \beta^2
\varepsilon}.
\end{equation}
Although this relation (see solid line in Fig.~\ref{fig5}c) gives
a reasonable estimate for the $\varepsilon$-dependence of $\tau$
it does not correctly reflect the empiric dependence. Apparently
this is a finite size effect, which is expected to become
significant at small values of $\varepsilon$. Roth et al.
\cite{roth94} recently demonstrated that the effective phase
diffusion constant, measurable in RBC and TCF relaxation
experiments, depends sensitively on the aspect ratio
$\alpha=\sqrt{\varepsilon}L/\xi_0$, defined by the quotient
between the container dimension and the linear correlation length.
Taking TCF as an example, a decrease of $\alpha$ from $50$ down
to $15$ (which in our experiment corresponds to a reduction of
$\varepsilon$ form $9\%$ to $2 \%$) implies a decay of the
relaxation time by $20-30\%$. This is of the same order of
magnitude as the value observed in our measurements (see
Fig.~\ref{fig5}c).
%


{\it Acknowledgements\/} ---  We thank M.~L\"{u}cke for helpful
comments and J.~Albers for his support. This work is financially
supported by the Deutsche Forschungsgemeinschaft.

%
%

%
%

\end{document}